\documentclass[aps,prb,twocolumn]{revtex4}
\usepackage{graphicx}
\usepackage{natbib}

\begin{document}

\title{Ion and polymer dynamics in polymer electrolytes PPO-LiClO$_4$: \\insights from NMR line-shape analysis}

\author{M.\ Vogel}
\author{T.\ Thorbr\"ugge}
\affiliation{Institut f\"ur Physikalische Chemie, Westf\"alische
Wilhelms-Universit\"at M\"unster, Corrensstr.\ 30/36, 48149
M\"unster, Germany}

\date{\today}

\begin{abstract}
We investigate ion and polymer dynamics in polymer electrolytes PPO-LiClO$_4$ performing
$^2$H and $^7$Li NMR line-shape analysis. Comparison of temperature dependent $^7$Li and
$^2$H NMR spectra gives evidence for a coupling of ion and polymer dynamics. $^2$H NMR
spectra for various salt concentrations reveal a strong slow down of the polymer
segmental motion when the salt content is increased. The $^2$H NMR line shape further
indicates that the segmental motion is governed by dynamical heterogeneities. While the
width of the distribution of correlation times $G(\log \tau)$ is moderate for low and
high salt content, an extremely broad distribution exists for an intermediate salt
concentration 15:1 PPO-LiClO$_4$. For the latter composition, a weighed superposition of
two spectral components, reflecting the fast and the slow polymer segments of the
distribution, describes the $^2$H NMR line shape over a broad temperature range. Analysis
of the temperature dependent relative intensity of both spectral components indicates the
existence of a continuous rather than a discontinuous distribution $G(\log \tau)$. Such
continuous distribution is consistent with gradual fluctuation of the local salt
concentration and, hence, of the local environments of the polymer segments, whereas it
is at variance with the existence of large salt-depleted and salt-rich domains featuring
fast and slow polymer dynamics, respectively. Finally, for all studied mixtures
PPO-LiClO$_4$, the $^2$H NMR line shape strongly depends on the echo delay in the applied
echo-pulse sequence, indicating that the structural relaxation of the polymer segments
involves successive rotational jumps about small angles $\gamma\!<\!20^{\circ}$.

\end{abstract}
\maketitle

\section{Introduction}

Many polymers can dissolve salts so as to form mixtures supporting high ionic
conductivity. These polymer electrolytes are promising candidates for applications in
electrochemical devices such as solid state batteries. However, the use of these
materials is limited by the achievable electric conductivities at ambient temperature
and, hence, there is considerable interest to accelerate the ionic diffusion. Such
efforts require a thorough understanding of the mechanisms for the ion transport. Despite
considerable progress over the past decades, a general theory of ion dynamics in polymer
electrolytes is still lacking.

The interactions of cations and anions with the polymer matrix are important factors
governing the charge transport in polymer electrolytes. It is well established that the
segmental motion of the host polymer facilitates the ionic diffusion.\cite{Armand,
Ratner} However, the understanding of this coupling is incomplete. For example, it was
emphasized that couplings for the ionic species may differ since cation-polymer
interactions are stronger than anion-polymer interactions.\cite{Vincent} In addition,
structural inhomogeneities can affect ionic diffusion in polymer electrolytes. Along
these lines, it was stressed that the spatial distribution of the ions, in particular the
formation of ionic aggregates, is a relevant factor of the charge transport in polymer
electrolytes.\cite{Vincent,Stolwijk,Ferry,Frech} In general, the relevance of all these
aspects depends on sample composition and temperature, leading to a further complication
for a comprehensive theoretical treatment.

Polymer electrolytes based on polypropylene oxide (PPO) have attracted much attention
because of their high ionic conductivities together with an amorphous nature.
\cite{Armand} We use $^2$H and $^7$Li NMR to investigate mixtures PPO-LiClO$_4$ with
different salt concentrations or, equivalently, different ether oxygen to lithium ratios,
O:Li. In the past, various experimental methods were used to characterize structure and
dynamics of the PPO-LiClO$_4$ system. The electric conductivity $\sigma_{dc}$ at room
temperature shows a broad maximum at intermediate O:Li ratios ranging from about 10:1 to
30:1.\cite{McLin,Furukawa} For such intermediate salt concentrations, differential
scanning calorimetry (DSC) revealed two glass transition
steps,\cite{Moacanin,Vachon_1,Vachon_2} suggesting a liquid-liquid phase separation into
salt-rich and salt-depleted regions. Likewise, an inhomogeneous structure in the
intermediate composition range was proposed to explain results from dielectric
spectroscopy (DS),\cite{Furukawa} which give evidence for the existence of two processes
related to fast and slow polymer dynamics. Consistently, for PPO-NaCF$_3$SO$_3$, fast and
slow relaxation processes were found using photon correlation spectroscopy
(PCS).\cite{Bergman} In contrast, other studies on polymer electrolytes PPO-LiClO$_4$
challenged the existence of salt-rich and salt-depleted microphases. Small angle neutron
scattering indicated that there are no structural inhomogeneities on a length scale of
several nanometers, \cite{Carlsson_1} in harmony with the interpretation of data from
coherent quasielastic neutron scattering (QENS).\cite{Carlsson_2} Finally, it was
proposed that these apparent discrepancies can be reconciled based on a structure model
resulting from reverse Monte-Carlo (RMC) simulations of diffraction
data.\cite{Carlsson_3} Specifically, the RMC model of PPO-LiClO$_4$ features salt-rich
and salt-depleted regions on a small length scale of about $\mathrm{1\,nm}$. In addition,
it reveals that the chain conformation changes as a consequence of coordination to the
cations.

In view of the importance of the polymer dynamics for the ionic migration, there is
considerable interest to ascertain how presence of salt affects the segmental motion in
polymer electrolytes. For PPO as host polymer, it was established that addition of salt
slows down the polymer dynamics.\cite{Furukawa,Bergman,Carlsson_2,Torell,Roux} This
effect was regarded as a consequence of cations acting as transient intra- or
intermolecular cross links between different monomers, leading to a reduction of chain
flexibility. Furthermore, neutron and light scattering studies reported that the PPO
dynamics is more nonexponential when salt is present.\cite{Carlsson_2,Torell} This
additional stretching was explained by either heterogeneous effects, \cite{Torell} i.e.,
a broadening of the distribution of correlation times $G(\log \tau)$, or homogenous
effects, \cite{Carlsson_2} i.e., intrinsic stretching. Finally, a DS study on
PPO-LiClO$_4$ reported that, over a broad concentration and temperature range, there is a
linear relationship between the ionic diffusion coefficient and the relaxation rate of
the polymer segments in contact with salt,\cite{Furukawa} suggesting that the segmental
motion triggers the elementary steps of the ionic migration.

Concerning NMR work on polymer electrolytes, several studies provided insights into the
ion dynamics, whereas applications on the polymer motion are rare. In detail, NMR
field-gradient techniques were used to determine cationic and anionic self diffusion
coefficients,\cite{Vincent,Arumugam,Ward} while measurements of the NMR spin-lattice
relaxation and the line shape provided information about ion dynamics on a more local
scale.\cite{Roux,Chung_1,Chung_2,Adamic, Fan,Donoso,Forsyth} On the other hand,
investigation of polymer dynamics was limited to analysis of $^1$H spin-lattice
relaxation at high temperatures.\cite{Ward,Roux,Donoso} Here, we use $^7$Li and $^2$H NMR
to study both the ion and the polymer dynamics in polymer electrolytes PPO-LiClO$_4$. In
particular, we exploit that $^2$H NMR provides straightforward access to the rotational
jumps of specific polymer segments.\cite{Spiess} While $^2$H NMR line-shape analysis is
in the focus of the present contribution, future work will present $^2$H NMR multi-time
correlation functions. Using this concerted approach, we ascertain in detail to which
extent the presence of salt affects the segmental motion in polymer electrolytes.

\section{Experiment}\label{exp}

In addition to the neat PPO melt, we investigate polymer electrolytes PPO-LiClO$_4$ with
three salt concentrations, namely, 30:1, 15:1 and 6:1. PPO
($M_w\!=\!\mathrm{6510\,g/mol}$) and LiClO$_4$ were purchased from Polymer Source Inc.\
and Aldrich, respectively. Strictly speaking, the used polymer was poly(propylene
glycol), having terminal OH groups. To prepare the polymer electrolytes, weighed amounts
of PPO and LiClO$_4$ were dissolved in acetonitrile. The resulting mixtures were stirred
at 80$^\circ$C until a clear homogeneous solution was obtained. Afterwards, solvent and
moisture were removed on a vacuum line for several days. The resulting samples were
sealed in the NMR tube. $^1$H and $^{13}$C NMR spectra confirmed absence of water and
solvent.

A Netsch DSC-200 instrument was used to perform DSC experiments at a heating rate of
$\mathrm{10\,K/min}$. We find glass transition temperatures $T_g\!=\!\mathrm{203\pm2\,K}$
for PPO and 30:1 PPO-LiClO$_4$, consistent with results reported in the
literature.\cite{Furukawa,McLin,Torell,Vachon_1} For 6:1 PPO-LiClO$_4$, we obtain
$T_g\!=\!\mathrm{272\pm2\,K}$, which is $10\!-\!20\,\mathrm{K}$ smaller than the
literature value,\cite{Vachon_1} suggesting some differences in the respective
compositions. For 15:1 PPO-LiClO$_4$, a very broad glass transition range extending from
about $\mathrm{205\,K}$ to $\mathrm{245\,K}$ is observed. In previous studies, two glass
transition temperatures $T_{g,1}\!\approx\!\mathrm{208\,K}$ and
$T_{g,2}\!\approx\!\mathrm{240\,K}$ were found for the 16:1 composition.\cite{Vachon_1}

The $^2$H ($^7$Li) NMR experiments were performed on Bruker DSX 400 and DSX 500
spectrometers working at Larmor frequencies $\omega_0/2\pi$ of $\mathrm{61.4\,MHz}$
($\mathrm{155.6\,MHz}$) and $\mathrm{76.8\,MHz}$ ($\mathrm{194.4\,MHz}$), respectively.
Two Bruker probes were used to apply the radio frequency pulses resulting in $90^\circ$
pulse lengths between $\mathrm{2.0\,\mu s}$ and $\mathrm{3.5\,\mu s}$, depending on the
nucleus and setup. A flow of nitrogen gas controlled by a Bruker VT 3000 heading unit was
utilized to adjust the sample temperature. To remove deviations between set and actual
temperature within an uncertainty of $\pm\mathrm{1.5\,K}$, temperature calibration was
done using $^{207}$Pb NMR spectra of lead nitrate.\cite{Takahashi} Comparing $^2$H NMR
spectra for different setups, we determined that neither the magnetic field strength nor
the probe affects the results. Moreover, changing the sample position within the coil, we
found no evidence for the presence of temperature gradients in the sample. The solid-echo
pulse sequence was applied to measure the $^2$H and $^7$Li NMR spectra. To take into
account the respective nuclear spin $I$,\cite{Kanert} the sequences $(90^\circ)_x$ -
$t_p$ - $(90^\circ)_y$ and $(90^\circ)_x$ - $t_p$ - $(64^\circ)_y$ were used in $^2$H NMR
and $^7$Li NMR, respectively. The $^2$H spin-lattice relaxation times were measured
utilizing the saturation-recovery pulse sequence.

\section{Theory}

\subsection{Basics of $^2$H and $^7$Li NMR}\label{NMR}

In solid-state $^2$H NMR and $^7$Li NMR, the first order quadrupolar interaction,
describing the interaction of the nuclear quadrupole moment with the electric field
gradient (EFG) at the nuclear site, is the dominant internal interaction. It results in a
frequency shift, $\omega_Q$, which is detected in the experiment. Specifically, the
quadrupolar precession frequency $\omega_Q$ is given by\cite{Spiess}
\begin{equation}\label{omega_1}
\omega_Q(\theta,\phi)=\pm \frac{\delta}{2}(3\cos^2\theta -1-\eta\sin^2\theta\cos2\phi)
\end{equation}
Here, $\theta$ and $\phi$ specify the orientation of the principal axes system of the EFG
tensor with respect to the external static magnetic flux density $\mathbf{B_0}$ and
$\eta$ is the asymmetry parameter of this tensor. The anisotropy parameter $\delta$ is
related to the quadrupolar coupling constant $C\!=\!e^2qQ/\hbar$ according to
$\delta\!=\!\frac{3}{4}\,C$ for $^2$H ($I\!=\!1$) and $\delta\!=\!\frac{1}{2}\,C$ for
$^7$Li ($I\!=\!3/2$). For the $^2$H nucleus, the two signs in Eq.\ (\ref{omega_1})
correspond to the two allowed transitions between the three Zeeman levels. For the $^7$Li
nucleus, these signs are associated with two satellite transitions
$\pm3/2\!\leftrightarrow\! \pm1/2$, while the first order quadrupolar interaction does
not affect the frequency of the central transition $1/2\!\leftrightarrow\! -1/2$.

In $^2$H NMR, the EFG tensor is intimately linked to the molecular frame. In our case of
deuterated PPO, [CD$_2$-CD(CD$_3$)-O]$_\mathrm{n}$, the monomeric unit features three
deuterons in the backbone and three deuterons in the methyl group, which will be denoted
as B deuterons and M deuterons, respectively. For the B deuterons, the EFG tensor is
axially symmetric, i.e., $\eta\!=\!0$, and its principal $z$ axis points along the
direction of the C-D bond.\cite{Spiess} For the M deuterons, fast rotation of the methyl
group leads to an averaging of the coupling tensor at the studied temperatures. The
averaged tensor is also axially symmetric and its principal $z$ axis is aligned with the
three-fold symmetry axis of the methyl group.\cite{Spiess} Altogether, the quadrupolar
precession frequency for the B and M deuterons can be written as
\begin{equation}\label{omega_2}
\omega_Q(\theta)=\pm \frac{\delta_{B,M}}{2}\,(3\cos^2\theta_{B,M} -1)
\end{equation}
Here, due to the motional averaging, the anisotropy parameters of the B and the M
deuterons are related according to $\delta_{B} \!=\!3\,\delta_{M}$.\cite{Spiess} For the
B deuterons, $\theta_{B}$ specifies the angle between $\mathbf{B_0}$ and the direction of
the C-D bond, whereas, for the M deuterons, $\theta_M$ denotes the angle between
$\mathbf{B_0}$ and the three-fold symmetry axis of the methyl group. In polymer melts,
there is an isotropic distribution of molecular orientations and, hence, a broad spectrum
results from the powder average at low temperatures, when molecular dynamics can be
neglected. In $^2$H NMR, this rigid-lattice spectrum has a specific line shape, called
Pake spectrum.\cite{Spiess} For PPO, we expect a superposition of two Pake spectra since
the anisotropy parameters for the B and the M deuterons are different.

In $^7$Li NMR, rigid-lattice spectra are comprised of two contributions resulting from
the central and the satellite transitions, respectively. The central transition, being
independent of the quadrupolar interaction, yields a narrow line, while a broad spectral
component results from the satellite transitions due to the action of the quadrupolar
interaction together with the powder average, see Eq.~(\ref{omega_1}). In polymer
electrolytes, the large diversity of lithium ionic environments leads to a broad variety
of EFG tensors and, hence, values of $\theta$ and $\eta$. Therefore, one expects a rather
unstructured satellite component.\cite{Chung_1}

\subsection{$^2$H and $^7$Li NMR line-shape analysis}\label{DYN}

Molecular dynamics with correlation times $\tau$ on the order of the inverse spectral
width, i.e., on the order of microseconds, results in a collapse of the discussed $^2$H
and $^7$Li NMR rigid-lattice spectra. In $^2$H NMR, there is a well defined relation
between the resonance frequency and the orientation of specific molecular units, see Eq.\
(\ref{omega_2}), and, hence, analysis of the time dependence $\omega_Q(t)$ provides
straightforward access to the polymer segmental motion. In $^7$Li NMR, line-narrowing of
the satellite component indicates a variation of the EFG tensor at the nuclear site with
time, which can result from either lithium ionic motion or, for a static ion, from a
rearrangement of the neighboring polymer chains. In polymer electrolytes, both dynamic
processes occur on a similar time scale and, hence, it is difficult to disentangle
effects due to ion and polymer dynamics. The central line is broadened due to dipolar
interactions between the magnetic moments of different nuclei, which are more important
in $^7$Li than in $^2$H NMR. Since homonuclear and heteronuclear dipole-dipole
interactions contribute, also the central line narrows as a consequence of both ion and
polymer dynamics.

\subsubsection{$^2$H NMR two-phase spectra}

To discuss effects of molecular dynamics on the $^2$H NMR line shape in more detail, let
us assume that there is a single anisotropy parameter $\delta$. Further, we suppose that
the polymer segments show isotropic reorientation, as may be expected in the
melt.\cite{Spiess} If this dynamics is characterized by a single correlation time $\tau$,
a rigid-lattice spectrum and a Lorentzian line will be found at low and high
temperatures, respectively, where the limits of slow motion ($\tau\!\gg\!1/\delta$) and
fast motion ($\tau\!\ll\!1/\delta$) are valid, while a crossover will occur for
$\tau\!\approx\!1/\delta$. For polymer electrolytes, a distribution of correlation times
can be expected so that both fast and slow segmental reorientation may occur at the same
temperature. Then, the spectrum would be comprised of a weighed superposition of a
rigid-lattice spectrum and a narrow line.\cite{Roessler_2,Roessler} In general, such
''two-phase spectra'' can result from two limiting cases for the shape of the
distribution of correlation times $G(\log \tau)$. On the one hand, it is possible that
the existence of well defined microphases is reflected in a discontinuous distribution of
correlation times, which is comprised of two narrow non-overlapping contributions
$G_f(\log \tau_f)$ and $G_s(\log \tau_s)$ associated with fast segmental motion in
salt-depleted domains ($\tau_f\!\!\ll\!1/\delta$) and slow segmental motion salt-rich
domains ($\tau_s\!\!\gg\!1/\delta$). On the other hand, one can imagine that gradual
fluctuations in the local salt concentration lead to a wide variety of local environments
that manifests themselves in a very broad continuous distribution $G(\log \tau)$. Then,
the fast and slow segments of this continuous distribution would yield a narrow line and
a rigid-lattice spectrum, respectively, whereas contributions from segments with
$\tau\!\approx\!1/\delta$ are negligible.\cite{Roessler_2,Roessler}

In both limiting cases, the temperature dependent normalized spectral intensity
$S(\omega;T)$ can be written as\cite{Roessler_2,Roessler}
\begin{equation}\label{ZP}
S(\omega;T)=W(T)S_f(\omega)+[1-W(T)]S_s(\omega)
\end{equation}
Here, $S_f(\omega)$ and $S_s(\omega)$ are the normalized line shapes for the limits of
fast and slow motion, respectively. The weighting factor $W(T)$ of the former spectral
pattern ($0\!\leq\!W\!\leq\!1$) is related with the distribution of correlation times
according to
\begin{equation}
W(T)=\int_{-\infty}^{\log 1/\delta} G(\log \tau;T)\,d\log \tau
\end{equation}

Based on the temperature dependence of the weighting factor, it is possible to
distinguish between both limiting cases for the shape of the distribution $G(\log \tau)$.
For a discontinuous distribution, two rapid rises of $W(T)$ separate three temperature
ranges in which the weighting factor is constant. Specifically, $W(T)\!=\!0$ at low
temperatures ($\tau_f,\tau_s\!\gg\!1/\delta$), $W(T)\!=\!W_f$ at intermediate
temperatures ($\tau_f\!\ll\!1/\delta\!\ll\!\tau_s$), and $W(T)\!=\!1$ at high
temperatures ($\tau_f,\tau_s\!\ll\!1/\delta$). Here, $W_f$ is the contribution of
$G_f(\log \tau_f)$ to the total distribution of correlation times. In contrast, the
temperarure dependent shift of a broad continuous $G(\log \tau)$ leads to a gradual
variation of $W(T)$. For a logarithmic Gaussian distribution, characterized by a
temperature independent width,
\begin{equation}\label{Gauss}
G(\log \tau; T)=\frac{1}{\sqrt{2\pi}\sigma}\exp\left(-\frac{[\,\log
\tau\!-\!\log\tau_m(T)]^2}{2\sigma^2}\right),
\end{equation}
one obtains after short calculation
\begin{equation}\label{W_T}
W(T)=\frac{1}{2}+\frac{1}{2}\,\mathrm{erf}[x(T)]
\end{equation}
where $x\!=\!(\log(1/\delta)\!-\!\log\tau_m)/(\sqrt 2\sigma)$ and $\mathrm{erf}(x)$ is
the error function. Therefore, given the temperature dependent shift of $G(\log \tau)$ is
known, analysis of $W(T)$ provides information about the width of this distribution, as
will be used below.

\begin{figure}
\includegraphics[angle=270,width=8.2cm]{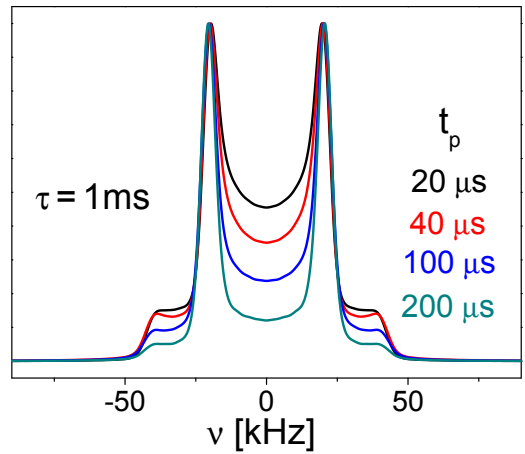}
\caption{(Color online) Simulated $^{2}$H NMR spectra for the indicated echo delays $t_p$
in the solid-echo pulse sequence. In the RW simulations, we assumed that C-D bonds
($\delta\!=\!2\pi\cdot\mathrm{40\,kHz}$, $\eta\!=\!0$) perform isotropic rotational jumps
about a jump angle $\gamma\!=\!15^{\circ}$. Further, we used a correlation time
$\tau\!=\!\mathrm{1\,ms}$, corresponding to a jump correlation time
$\tau_j\!\approx\!\mathrm{100\,\mu s}$, see Eq.\ (\ref{gamma}).}\label{RW}
\end{figure}

\subsubsection{$^2$H NMR spectra for various solid-echo delays}

Finally, we illustrate that, in $^2$H NMR, variation of the echo delay $t_p$ in the
solid-echo pulse sequence provides access to the reorientational mechanism, as was shown
for various types of
motion.\cite{Spiess_2,Vold,Jones,Pschorn,Vogel_1,Vogel_2,Roessler,Vogel_3,Vogel_4,Aze}
For polymer melts, it was reported that the isotropic reorientation of the polymer
segments during the $\alpha$ process results from successive rotational jumps about small
angles.\cite{Spiess,Pschorn,Tracht} Then, the elementary rotational jumps occur on a much
shorter time scale, than the overall loss of correlation. Strictly speaking, the ratio of
the jump correlation time $\tau_j$ and the correlation time $\tau$ depends on the jump
angle $\gamma$ according to\cite{Anderson}
\begin{equation}\label{gamma}
\frac{\tau_j}{\tau}=\frac{3}{2}\sin^2\gamma.
\end{equation}
Thus, for small jump angles $\gamma\!<\!20^{\circ}$, say, one may encounter a situation
$\tau_j\approx\!t_p\!\ll\!\tau$ at appropriate temperatures. In the literature, it was
shown that, in this case, the $^2$H NMR line shape shows a characteristic dependence on
the echo delay $t_p$. \cite{Pschorn,Vogel_1,Roessler,Vogel_3} To illustrate the effects,
we calculate $^2$H NMR spectra from random-walk (RW) computer simulations. The
methodology of these computation was described in previous work.\cite{Vogel_1,Vogel_3}
Here, we assume that C-D bonds perform isotropic rotational jumps with a jump angle
$\gamma\!=\!15^{\circ}$, which is a typical value for the $\alpha$ process near the glass
transition.\cite{Tracht,Roessler} Further, we use a correlation time
$\tau\!=\!\mathrm{1\,ms}$, corresponding to a jump correlation time
$\tau_j\!\approx\!\mathrm{100\,\mu s}$. In Fig.\ \ref{RW}, we show $^2$H NMR spectra
resulting from such dynamics for various values of $t_p$. We see that the intensity in
the center of the solid-echo spectrum strongly decreases with respect to the intensity of
the ''horns'' when the echo delay is extended. The characteristics of this effect depend
on the value of the correlation time $\tau$ and the jump angle $\gamma$ so that, in
principle, the $t_p$ dependence of the line shape enables determination of these
quantities. However, we refrain from such analysis since unambiguous quantification will
be difficult if distributions of correlation times and jump angles exist, as expected in
our case.

\section{Results}

\begin{figure}
\includegraphics[angle=270,width=8.2cm]{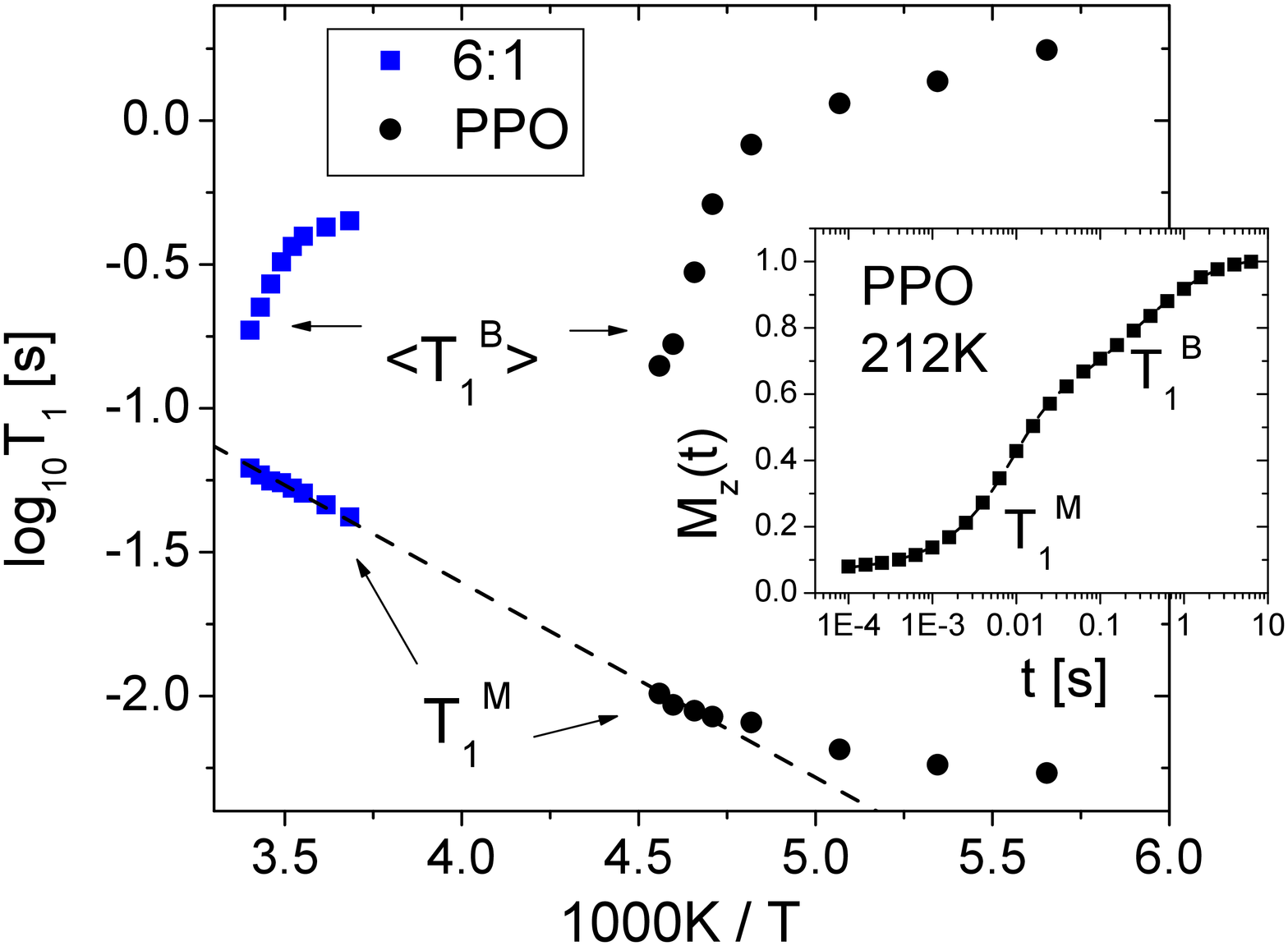}
\caption{(Color online) Temperature dependent $^{2}$H NMR spin-lattice relaxation times
for PPO and 6:1 PPO-LiClO$_4$. The measurements were performed at a Larmor frequency
$\omega_0/(2\pi)\!=\!\mathrm{61.4\,MHz}$. $T_1^{\,M}$ denotes the relaxation time of the
deuterons in the methyl group and $\langle T_1^{\,B} \rangle$ is the mean spin-lattice
relaxation time of the deuterons in the backbone. The dashed line is an Arrhenius fit to
the data $T_1^{\,M}$ at $T\!>\mathrm{200\,K}$. The inset shows the recovery of the
normalized longitudinal magnetization after saturation for PPO at $\mathrm{222\,K}$. The
dashed line is a fit to Eq.\ (\ref{FIT}). }\label{T1}
\end{figure}

\subsection{$^2$H NMR spin-lattice relaxation}

First, we measure the $^2$H NMR spin-lattice relaxation of the mixtures PPO-LiClO$_4$.
The inset of Fig.\ \ref{T1} shows the recovery of the normalized longitudinal
magnetization $M_z(t)$ for PPO at $T\!=\!\mathrm{222\,K}$. We see a two-step increase as
a consequence of diverse spin-lattice relaxation behaviors of the different deuteron
species. Specifically, due to fast three-fold jumps of the methyl group, the M deuterons
exhibit faster spin-lattice relaxation than the B deuterons. Therefore, we interpolate
the data with the function
\begin{equation}\label{FIT}
M_z(t)\!=\!1\!-\!a_M\exp\left(-\frac{t}{T_1^{\,M}}\right)\!-\!a_B\exp\left[-\left(\frac{t}{T_1^{\,B}}\right)^\beta\right]
\end{equation}
to extract the respective spin-lattice relaxation times. In doing so, we consider our
finding that the spin-lattice relaxation of the M deuterons ($T_1^{\,M}$) and the B
deuterons ($T_1^{\,B}$) is exponential and nonexponential, respectively, see below. In
harmony with an equal amount of M and B deuterons in the monomeric unit of the studied
polymer, we obtain coefficients $a_M\!\approx\!a_B\!\approx\!0.5$ from these fits. In
Fig.\ \ref{T1}, we see that the spin-lattice relaxation time $T_1^{\,M}$ decreases upon
cooling, indicating $\omega_0\tau\!\ll\!1$ for the methyl group jumps at the studied
temperatures.\cite{BPP} Below $\mathrm{200\,K}$, a somewhat weaker temperature dependence
of $T_1^{\,M}$ suggests that the $T_1$ minimum ($\omega_0\tau\!\approx\!1$) is
approached. Above $\mathrm{200\,K}$, the values $T_1^{\,M}$ for PPO and 6:1 PPO-LiClO$_4$
fall on the same line and, hence, the presence of salt does not affect the methyl group
motion. Interpolation of these data with an Arrhenius law yields an activation energy
$E_a\!=\!\mathrm{13\,kJ/mol}$. Consistent with our results, QENS work on PPO-LiClO$_4$
found that an exponential correlation function, which is not affected by addition of
salt, describes the methyl group jumps.\cite{Ea,Andersson} However, this study reported a
somewhat higher activation energy of $\mathrm{17\,kJ/mol}$.

In harmony with previous $^2$H NMR work on molecular and polymeric glass formers near
$T_g$,\cite{Roessler,Schnauss,Leissen} we find that the $^2$H spin-lattice relaxation of
the B deuterons is nonexponential. Fitting the data for PPO at $\mathrm{222\,K}$ to
Eq.~(\ref{FIT}), we obtain a stretching parameter $\beta\!=\!0.48$, see Fig.\ \ref{T1}.
For the studied polymer electrolytes PPO-LiClO$_4$, including the compositions 15:1 and
30:1, which were measured at a Larmor frequency $\omega_0/(2\pi)\!=\!\mathrm{76.8\,MHz}$,
we find stretching parameters in the range $0.4\!\leq\beta\!\leq\!0.6$ at temperatures
somewhat above $T_g$. Therefore, we use the $\Gamma$-function to calculate the mean
spin-lattice relaxation time according to $\langle
T_1^{\,B}\rangle\!=\!(T_1^{\,B}/\beta)\;\Gamma(1/\beta)$. In Fig.\ \ref{T1}, it is
evident that $\langle T_1^{\,B} \rangle$ increases when the temperature is decreased,
indicating $\omega_0\tau\!\gg\!1$ for the segmental motion at the studied temperatures.
Further, we see that $\langle T_1^{\,B} (T)\rangle$ is shifted to higher temperatures for
6:1 PPO-LiClO$_4$, implying that addition of salt slows down the dynamics of the B
deuterons. In particular, for both samples, the curves $\langle T_1^{\,B} (T)\rangle$
show a kink near the respective glass transition temperature, as was observed for other
glass-forming liquids.\cite{Roessler,Lusceac}

In the literature,\cite{Roessler,Leissen} it was demonstrated that nonexponential $^2$H
spin-lattice relaxation is due to a distribution of spin-lattice relaxation times,
$V(T_1)$. Since $T_1$ depends on the spectral density $J(\omega)$, which is given by the
Fourier transform of the correlation function, such distribution $V(T_1)$ indicates the
existence of a distribution of correlation times. In general, fast relaxation processes,
e.g., a $\beta$ process, can dominate the spin-lattice relaxation near $T_g$ so that a
priori it is not clear whether this distribution of correlation times is associated with
the $\alpha$ process.\cite{Roessler} For PPO, findings in DS and PCS give no evidence for
the presence of fast processes other than the methyl group motion,\cite{Furukawa,Bergman}
consistent with our $^2$H NMR results, see below. Therefore, we conclude that the
nonexponential $^2$H spin-lattice relaxation of the B deuterons results from a
distribution $G(\log \tau)$ for the $\alpha$ process. Thus, dynamical heterogeneities
govern the structural relaxation of the host polymer in polymer electrolytes
PPO-LiClO$_4$, as was shown for neat polymer melts and supercooled molecular
liquids.\cite{Spiess,Roessler,13}

\subsection{$^2$H NMR line shape}\label{Specs}

\begin{figure}
\includegraphics[angle=0,width=7.5cm]{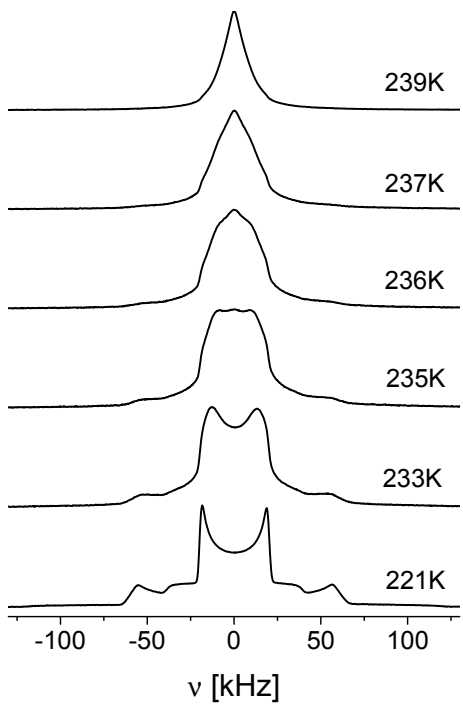} \caption{Temperature dependent $^{2}$H NMR
solid-echo spectra of PPO ($t_p\!=\!\mathrm{20\,\mu s}$).}\label{PPO}
\end{figure}

\begin{figure}
\includegraphics[angle=0,width=7.5cm]{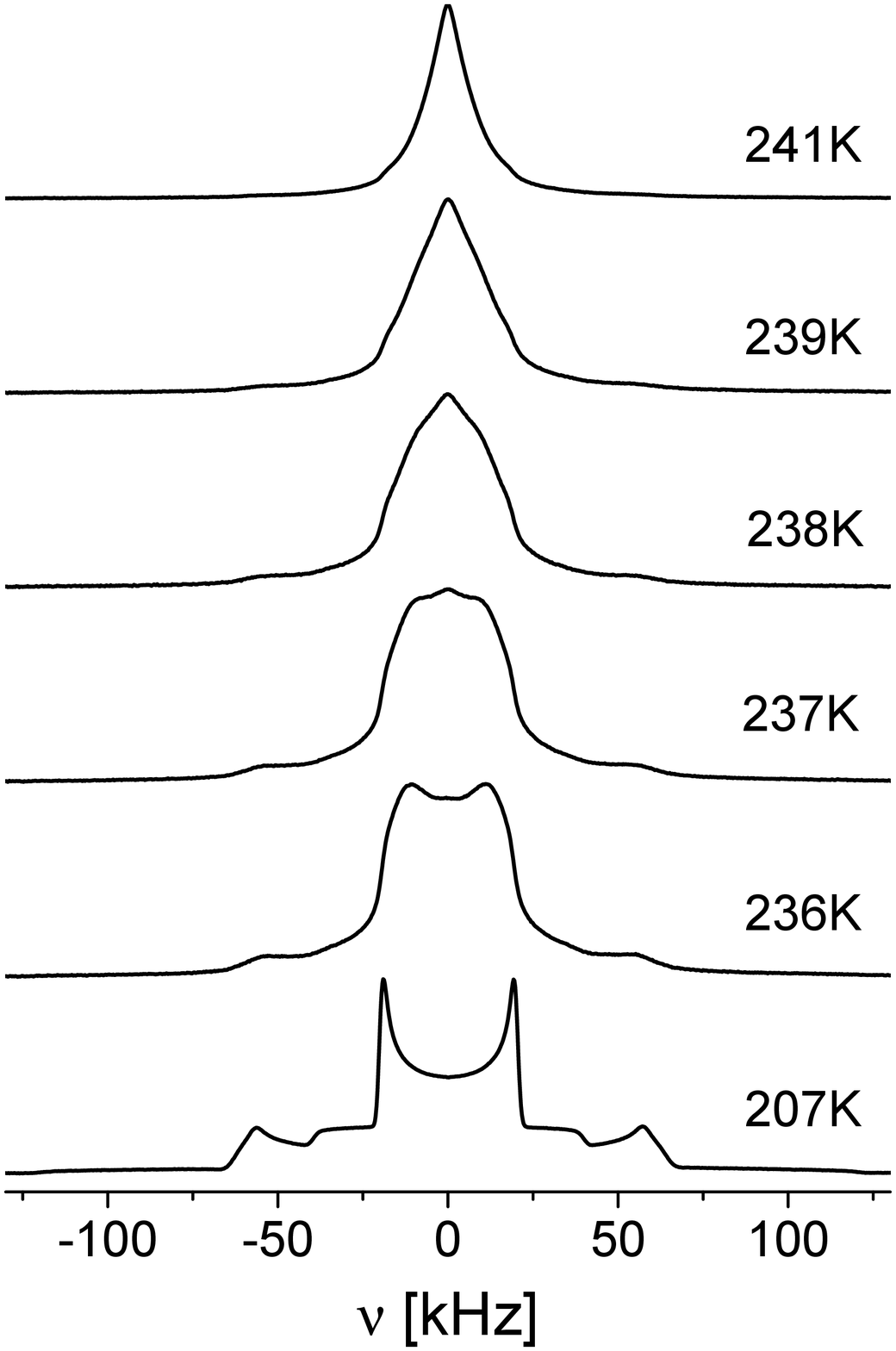}
\caption{Temperature dependent $^{2}$H NMR solid-echo spectra of 30:1 PPO-LiClO$_4$
($t_p\!=\!\mathrm{20\,\mu s}$).}\label{PPO30}
\end{figure}

\begin{figure}
\includegraphics[angle=0,width=7.5cm]{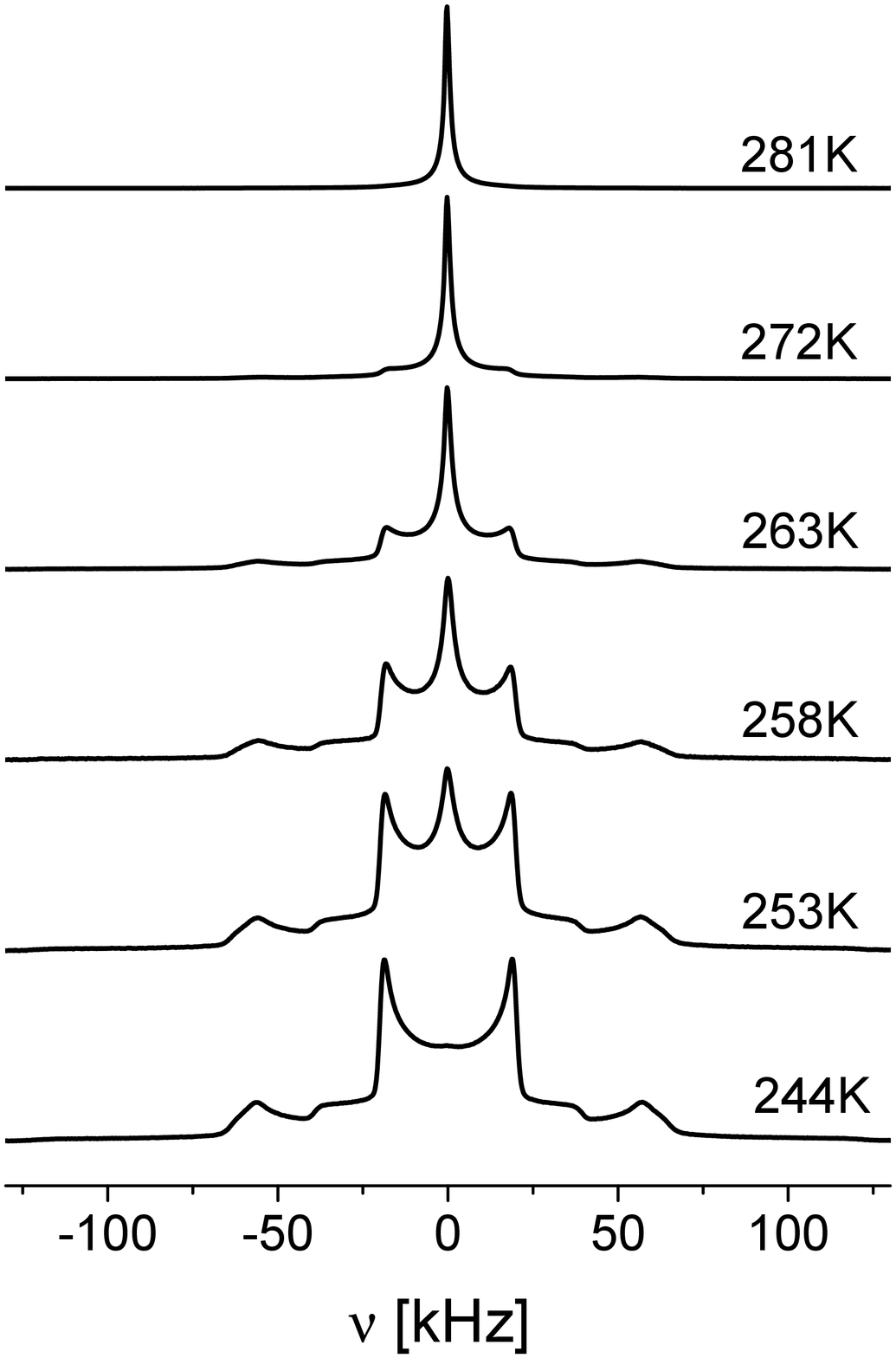}
\caption{Temperature dependent $^{2}$H NMR solid-echo spectra of 15:1 PPO-LiClO$_4$
($t_p\!=\!\mathrm{20\,\mu s}$).}\label{PPO15}
\end{figure}

\begin{figure}
\includegraphics[angle=0,width=7.5cm]{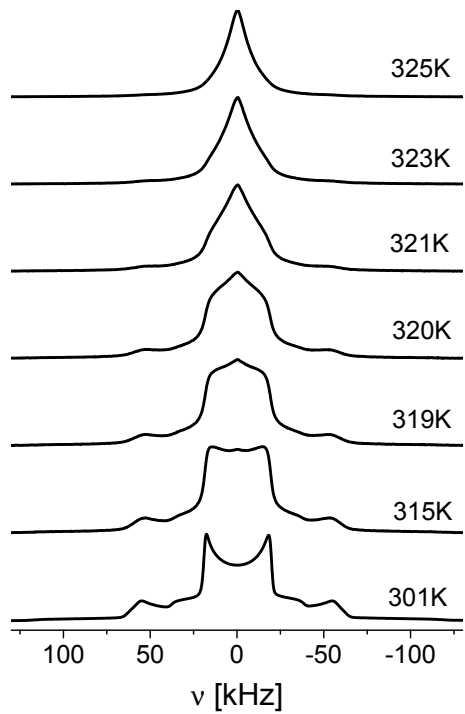}
\caption{Temperature dependent $^{2}$H NMR solid-echo spectra of 6:1 PPO-LiClO$_4$
($t_p\!=\!\mathrm{20\,\mu s}$).}\label{PPO6}
\end{figure}

In Figs.\ \ref{PPO}-\ref{PPO6}, we present temperature dependent $^2$H NMR spectra of
PPO-LiClO$_4$ mixtures. At low temperatures, the rigid-lattice spectra
($\tau\!\gg\!1/\delta$) are comprised of two Pake spectra, reflecting the two deuteron
species, see Sec.\ \ref{NMR}. Fitting the line shapes, we obtain anisotropy parameters
$\delta_B\!\approx\!\mathrm{118\!-\!121\,kHz}$ and
$\delta_M\!\approx\!\mathrm{38\!-\!40\,kHz}$. For both deuteron species, the lower values
of the specified line-width ranges are found for higher salt contents. At high
temperatures, we observe a narrow line, indicative of fast ($\tau\!\ll\!1/\delta$)
isotropic reorientation of the polymer segments. When we compare the results for the
different compositions, it is evident that the line shape changes at higher temperatures
for higher salt concentrations, indicating that the presence of salt slows down the
segmental motion. The temperatures, at which $\tau\!\approx\!1/\delta$, are about
$\mathrm{85\,K}$ higher for 6:1 PPO-LiClO$_4$ than for PPO, showing that the effects are
strong, cf.\ Figs.\ \ref{PPO} and \ref{PPO6}.

Closer inspection of the $^2$H NMR line-shape changes at intermediate temperatures
reveals interesting aspects of the polymer segmental motion in the PPO-LiClO$_4$
mixtures. For PPO, there is a more or less continuous collapse of the rigid-lattice
spectrum when the temperature is increased, see Fig.\ \ref{PPO}, consistent with previous
results on glass-forming liquids.\cite{Lusceac,Roessler} Revisiting Figs.\ \ref{PPO30}
and \ref{PPO6}, we see some deviations from a continuous line narrowing for the low
(30:1) and the high (6:1) salt concentrations. In narrow temperature ranges ($\Delta
T\!<\!\mathrm{10\,K}$), the $^2$H NMR line shape resembles a two-phase spectrum, see
Sec.\ \ref{DYN}, although the effect is not pronounced. In Fig.\ \ref{PPO15}, it is
obvious that the two-phase signature is much better resolved for the intermediate 15:1
salt concentration. Over a broad temperature range, the $^2$H NMR spectrum can be
described as a weighed superposition of a narrow line and a rigid-lattice spectrum,
indicating the coexistence of fast ($\tau\!\ll\!1/\delta$) and slow
($\tau\!\gg\!1/\delta$) polymer segmental motion. Hence, for all studied polymer
electrolytes PPO-LiClO$_4$, the $^2$H NMR line shape indicates that the segmental motion
exhibits dynamical heterogeneities. However, the degree of the nonuniformity of the
motion depends on the composition. While the dynamical heterogeneities are very prominent
for intermediate salt concentrations, the effect is weaker for both low and high salt
content.

\begin{figure}
\includegraphics[angle=270,width=8.2cm]{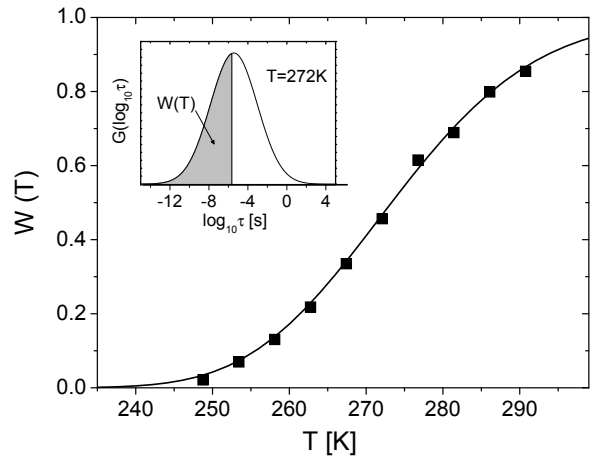}
\caption{Weighting factor $W(T)$ quantifying the relative intensity of the narrow line in
the two-phase spectra of 15:1 PPO-LiClO$_4$, see Fig.\ \ref{PPO15}. The data indicates
the existence of a broad continuous distribution $G(\log \tau)$. The solid line is an
expectation for $W(T)$ resulting from the assumption of a logarithmic Gaussian
distribution $G(\log_{10} \tau)$ characterized by a width parameter
$\sigma_{10}\!=\!2.4$, see text for details. The inset shows this distribution for
$T\!=\!272\mathrm{\,K}$. The shaded area marks the part of the distribution contributing
to the narrow line.}\label{WT}
\end{figure}

For a more detailed analysis of the two-phase spectra for 15:1 PPO-LiClO$_4$, we
determine the weighting factor $W(T)$ characterizing the relative contribution of the
narrow line, cf.\ Eq.\ (\ref{ZP}). In Fig.\ \ref{WT}, we see that $W(T)$ increases
gradually from $W\!=\!0$ to $W\!=\!1$ when the temperature is increased. As was discussed
in Sec.\ \ref{DYN}, such behavior indicates the existence of a continuous rather than a
discontinuous distribution of correlation times $G(\log \tau)$. Specifically, the latter
scenario would manifest itself in a discontinuous variation of the weighting factor.

To estimate the width of the distribution $G(\log \tau)$ for 15:1 PPO-LiClO$_4$, we use
input from our $^2$H NMR stimulated-echo study on this material.\cite{UP} There, we find
that the segmental motion exhibits nonexponential correlation functions, which can be
described by a Kohlrausch function, $\exp[-(t/\tau)^{\beta}]$. The small stretching
parameter $\beta\!\approx\!0.2$ shows no systematic variation with temperature,
consistent with a temperature independent width of the distribution $G(\log \tau)$. In
the temperature range of the two-phase signature, the mean correlation time follows an
Arrhenius law, $\log_{10}\langle\tau(T)\rangle\!=\!\log_{10} \tau_0+E_a/(T\ln 10)$, where
$\log_{10}\tau_0/\mathrm{s}\!=\!-49.3$ and $E_a\!=\!28100\mathrm{\,K}$. Based on these
results, we assume that the segmental motion exhibits a logarithmic Gaussian distribution
$G(\log_{10} \tau)$ that shifts with an activation energy $E_a\!=\!28100\mathrm{\,K}$,
where the width parameter $\sigma_{10}$ is temperature independent. For various values of
$\sigma_{10}$, we then calculate $W(T)$ according to Eq.\ (\ref{W_T}). In doing so, we
also vary the prefactor $\log_{10}\tau_0$ since, for broad logarithmic Gaussian
distributions, the mean correlation time $\langle\tau\rangle$ is longer than
$\tau_m$.\cite{PCCP} Further, we use $\log_{10} 1/\delta\!=\!-5.65$, which is the average
of the values resulting from $\delta_M$ and $\delta_B$, respectively.

In Fig.\ \ref{WT}, we see that the experimental data is well reproduced for a width
parameter $\sigma_{10}\!=\!2.4$ and a prefactor $\log_{10}\tau_0\!=\!-50.3$. This width
parameter $\sigma_{10}$ corresponds to a full width at half maximum of more than five
orders of magnitude and, hence, the polymer segmental motion in 15:1 PPO-LiClO$_4$ is
extremely heterogeneous, see inset of Fig.\ \ref{WT}. For this intermediate salt
concentration, the two-phase signature is observed in the temperature range
$240\mathrm{\,K}\lesssim\! T\!\lesssim\!310\mathrm{\,K}$. The lower and the higher end of
this range correspond roughly to the temperatures where the line-shape changes for PPO
and 6:1 PPO-LiClO$_4$ are found, see Figs.\ \ref{PPO} and \ref{PPO6}, suggesting that the
broad distribution of correlation times for the 15:1 composition results from large
fluctuations in the local salt concentration. Although the present results rule out the
presence of a discontinuous distribution $G(\log \tau)$ for the 15:1 composition, it is
possible that the broad continuous distribution is comprised of two distributions, which
are associated with polymer dynamics in salt-rich and salt-depleted regions, provided
these distributions overlap in large parts.

\begin{figure}
\includegraphics[angle=270,width=8.2cm]{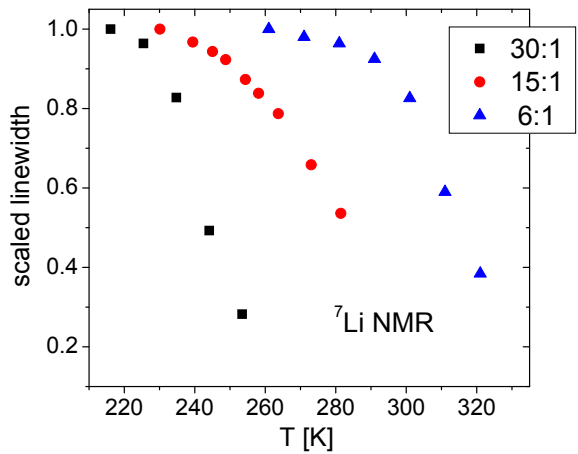}
\caption{(Color online) Full width at half height of the central transition in $^7$Li NMR
spectra of 30:1 PPO-LiClO$_4$, 15:1 PPO-LiClO$_4$ and 6:1 PPO-LiClO$_4$. For the sake of
comparison, the data was scaled by the value for the lowest temperatures studied.
}\label{7LI}
\end{figure}

\subsection{$^7$Li NMR line shape}

Next, we analyze $^7$Li NMR spectra to investigate the lithium ionic dynamics in
PPO-LiClO$_4$. For all studied polymer electrolytes, we find that the $^7$Li NMR
rigid-lattice spectra at low temperatures are comprised of a narrow and a broad Gaussian
associated with the central and the satellite transitions, respectively, see Sec.\
\ref{NMR}. The full width at half maximum (FWHM) of the satellite component amounts to
$\mathrm{83\,kHz}$ for the 6:1 composition and to about $\mathrm{42\,kHz}$ for the 15:1
and 30:1 compositions. Thus, the EFG tenors at the nuclear sites show a broader
distribution in shape for the high salt concentration. For the FWHM of the central
component, we find values of $\mathrm{5.1\,kHz}$ (30:1), $\mathrm{4.4\,kHz}$ (15:1) and
$\mathrm{2.3\,kHz}$ (6:1). Hence, an increase of the lithium ionic concentration is
accompanied by a line narrowing rather than a line broadening, suggesting that the
heteronuclear dipole-dipole interaction dominates the width of the central line.

In Fig.\ \ref{7LI}, we show the temperature dependent FWHM of the central line for the
30:1, 15:1 and 6:1 compositions. For the sake of comparison, the data were scaled to the
value for the respective rigid-lattice spectrum. Obviously, the line narrowing sets in at
higher temperatures for high salt concentrations, implying a slow down of the lithium
dynamics when the salt content is increased. For all compositions, rigid-lattice spectra
are observed up to $T_g$, indicating that there is no lithium ionic diffusion on the $\mu
s$-time scale in a rigid polymer matrix. A strong temperature dependence of the $^7$Li
line width is found at temperatures where $^2$H NMR line-shape changes are observed, see
Figs.\ \ref{PPO30}-\ref{PPO6}, suggesting a strong coupling of the lithium ionic and the
polymer segmental motion. However, one has to consider that both ion and polymer dynamics
render the $^7$Li NMR frequency time dependent and, hence, lead to a line narrowing so
that the $^7$Li line width does not necessarily provide access to the lithium ionic
diffusion. At $T\!>\!T_g\!+\!50\mathrm{\,K}$, the $^7$Li NMR spectra start to split up
into several lines, suggesting the existence of different lithium ionic environments.
Therefore, we refrain from analysis of the line width at higher temperatures.

\subsection{$^2$H NMR solid-echo spectra for various echo-delays}

\begin{figure}
\includegraphics[angle=270,width=8.2cm]{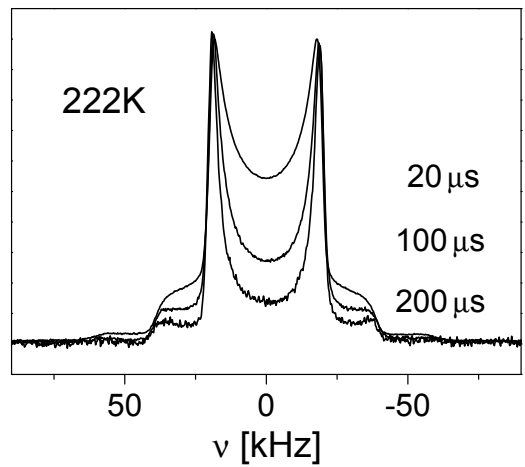}
\caption{Partially relaxed $^{2}$H NMR solid-echo spectra of PPO for the indicated
echo-delays $t_p$ and $T\!=\!\mathrm{222\,K}$. The delay between the saturation of the
longitudinal magnetization and the start of the solid-echo pulse sequence was
$\mathrm{3\,ms}$.}\label{TP}
\end{figure}

\begin{figure}
\includegraphics[angle=270,width=8.2cm]{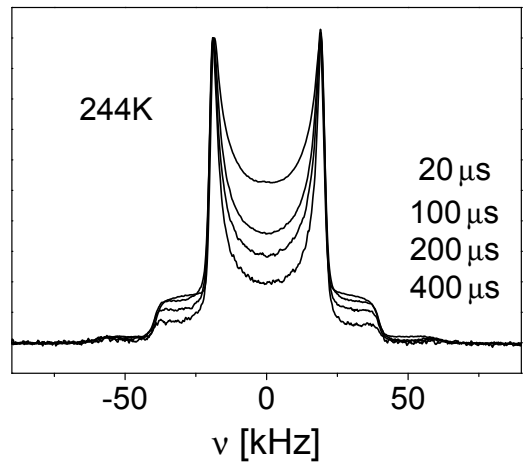}
\caption{Partially relaxed $^{2}$H NMR solid-echo spectra of 15:1 PPO-LiClO$_4$ for the
indicated echo-delays $t_p$ and $T\!=\!\mathrm{244\,K}$. The delay between the saturation
of the longitudinal magnetization and the start of the solid-echo pulse sequence was
$\mathrm{3\,ms}$.}\label{TP15}
\end{figure}

\begin{figure}
\includegraphics[angle=270,width=8.2cm]{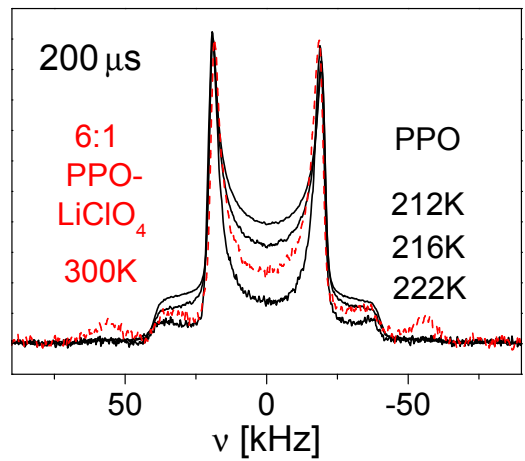}
\caption{(Color online) $^{2}$H NMR solid-echo spectra of PPO for the indicated
temperatures (solid lines) together with the $^{2}$H NMR solid-echo spectrum of 6:1
PPO-LiClO$_4$ for $T\!=\!\mathrm{300\,K}$ (dashed line). All spectra were measured using
an echo-delay $t_p\!=\!200\,\mu s$. The delay between the saturation of the longitudinal
magnetization and the start of the solid-echo pulse sequence was
$\mathrm{3\,ms}$.}\label{TPT}
\end{figure}

To study the mechanisms for the polymer segmental motion in the PPO-LiClO$_4$ systems, we
analyze the dependence of the $^2$H NMR solid-echo spectra on the echo delay $t_p$. In
our case, the superposition of two Pake spectra complicates such analysis. Therefore, we
exploit that the spin-lattice relaxation of the M deuterons is faster than that of the B
deuterons, see Fig.\ \ref{T1}, and single out the contribution of the former species in
partially relaxed $^2$H NMR spectra. In detail, after saturation of the longitudinal
magnetization, we do not wait for its complete recovery, but start the solid-echo pulse
sequence at times $t \approx \!T_1^{\,M}\!\ll\!T_1^{\,B}$, when the longitudinal
magnetization is dominated by the contribution of the M deuterons. In Figs.\
\ref{TP}-\ref{TPT}, we show partially relaxed $^2$H NMR solid-echo spectra of
PPO-LiClO$_4$ systems. Evidently, the narrow Pake spectrum of the M deuterons dominates
the line shape, whereas contributions from the B deuterons are suppressed in large parts.
Only for the 6:1 composition, suppression of the latter contribution is somewhat less
effective because of a smaller difference of the spin-lattice relaxation times $T_1^M$
and $T_1^B$, see Fig.\ \ref{T1}.

The dependence of the partially relaxed $^2$H NMR spectra on the solid-echo delay $t_p$
is displayed for PPO at $T\!=\!\mathrm{222\,K}$ in Fig.\ \ref{TP} and for 15:1
PPO-LiClO$_4$ at $T\!=\!\mathrm{244\,K}$ in Fig.\ \ref{TP15}. In both cases, a
rigid-lattice spectrum is observed for $t_p\!=\!20\,\mu s$, while the relative intensity
in the center of the spectrum decreases when the echo delay is extended. These line-shape
changes are a fingerprint of small angle rotational jumps, see Fig.\
\ref{RW}.\cite{Pschorn,Vogel_1,Vogel_2,Vogel_3} The specific effects can be explained
when we recall two effects, see Sec.\ \ref{DYN}. First, the angular resolution of the
experiment depends on the value of $t_p$ so that the line shapes for short and long echo
delays probe large-angle and small-angle displacements, respectively. Second, for
isotropic rotational jumps about small angles $\gamma\!\leq\!20^{\circ}$, say, the jump
correlation time $\tau_j$ is much shorter than the correlation time $\tau$, see Eq.\
(\ref{gamma}), since many elementary rotational jumps are required until the memory of
the initial orientation is lost completely. Thus, for such type of motion one encounters
the situation $\tau_j\!\approx\!t_p\!\ll\!\tau$ at appropriate temperatures. Then, the
elementary jumps lead to line-shape changes for large $t_p$, i.e., high angular
resolution, while a rigid-lattice spectrum results for short $t_p$, since angular
displacements, which are sufficiently large to as to lead to effects for poor spatial
resolution, are not achieved on the time scale of the experiment.

According to this argumentation, the $t_p$ dependence of the $^2$H NMR solid-echo spectra
in Figs.\ \ref{TP} and \ref{TP15} indicates that $\tau_j\!\approx\!t_p\!\ll\!\tau$ holds
for the reorientation of the threefold methyl group axes in PPO and 15:1 PPO-LiClO$_4$ at
the studied temperatures, consistent with the results Figs.\ \ref{PPO} and \ref{PPO15}.
This time scale separation $\tau_j\!\ll\!\tau$ is indicative of reorientation resulting
from small angle jumps, see Eq.\ (\ref{gamma}). For PPO, this reorientational mechanism
is consistent with results from previous work on polymer
melts.\cite{Spiess,Tracht,Pschorn} Our findings for 15:1 PPO-LiClO$_4$ together with that
for 6:1 PPO-LiClO$_4$, see Fig.\ \ref{TPT}, show that the isotropic reorientation of the
polymer segments in polymer electrolytes also involves successive small-angle jumps.
Hence, on a qualitative level, we find no evidence that addition of salt leads to a
change of the mechanism for the reorientation of the polymer segments. In a future
publication, we will show that analysis of $^2$H NMR two-time correlation functions
confirms this conclusion on a quantitative level.

Figure \ref{TPT} displays temperature dependent $^2$H NMR spectra of PPO for a large
solid-echo delay $t_p\!=\!200\,\mu s$. We see that the intensity in the center of the
spectrum is diminished for $T\!=\!222\mathrm{\,K}$, cf.\ also Fig.\ \ref{TP}, while this
effect becomes less prominent when the temperature is decreased, until the line shape is
independent of $t_p$ for $T\!\leq\!212\mathrm{\,K}$. Since the line shape for
$t_p\!=\!200\,\mu s$ is sensitive to small-angle motion, these results indicate that
cooling slows down the elementary rotational jumps so that, near $T_g$, any polymer
segmental motion is frozen in on the $\mu s$ time scale. We further emphasize that the
observed temperature dependence of the spectra for a large echo delay together with a
$t_p$ independent line shape at low temperatures rule out that the observed effects are
due to experimental imperfections, e.g., due to an inaccurate time origin for the Fourier
transformation of the time signal. Moreover, these observations exclude spin diffusion,
i.e., a transfer of magnetization as a consequence of flip-flop processes of the spins,
as possible origin of the $t_p$ dependent line shape. Since spin diffusion is not
directly related to molecular dynamics, it shows, if at all, a weak temperature
dependence and, hence, it cannot account for the pronounced effects in Fig.\
\ref{TPT}.\cite{Vogel_2}

Many molecular and polymeric glass formers show a secondary relaxation ($\beta$ process)
that is intrinsic to the glassy state.\cite{JG} In the
literature,\cite{Vogel_2,Vogel_3,Vogel_4} it was demonstrated that, at
$T\!\lesssim\!T_g$, the spatially restricted motion associated with the $\beta$ process
leads to a $t_p$ dependence of the $^2$H NMR solid-echo spectrum, which resembles that
resulting from the elementary steps of the $\alpha$ process at $T\!>\!T_g$, see Figs.\
\ref{TP}-\ref{TPT}. In previous NMR work, it was suggested that a $\beta$ process exists
in polymer electrolytes PPO-LiClO$_4$, too.\cite{Chung_2} Here, the $^2$H NMR solid-echo
spectra give no evidence that PPO-LiClO$_4$ mixtures electrolytes exhibit a $\beta$
process. In particular, we determined that the $^2$H NMR spectra of 15:1 PPO-LiClO$_4$
are independent of the solid-echo delay in the temperature range
$160\mathrm{\,K}\!\leq\!T\!\leq\!T_g$. Consistently, no $\beta$ process was oberved in DS
on polymer electrolytes PPO-LiClO$_4$.\cite{Furukawa}

\section{Conclusion and Summary}\label{Discussion}

Previous work on the polymer segmental motion in polymer electrolytes PPO-LiClO$_4$
reported nonexponential two-time correlation functions and broad dielectric loss
spectra.\cite{Furukawa,Carlsson_2} Two limiting scenarios can explain such nonexponential
relaxation. In the heterogeneous limit, all particles obey exponential correlation
functions, however, a distribution of correlation times $G(\log \tau)$ exists. In the
homogeneous limit, all particles obey the same correlation function, which is, however,
intrinsically non-exponential. For disordered materials, heterogeneous and homogeneous
contributions may coexist. Measurements of two-time correlation functions and, hence, the
previous approaches do not allow one to discriminate between both scenarios and to
quantify the relevance of homogeneous and heterogeneous contributions.\cite{13} Here, we
demonstrated that $^2$H NMR spin-lattice relaxation and line shape consistently indicate
that pronounced dynamical heterogeneities govern the polymer segmental motion in polymer
electrolytes PPO-LiClO$_4$. Specifically, at appropriate temperatures, the existence of a
distribution of correlation times $G(\log \tau)$ manifests itself in a nonexponential
$^2$H spin-lattice relaxation and in a $^2$H NMR two-phase line shape, i.e., the spectrum
is comprised of a weighed superposition of a narrow line and a rigid-lattice spectrum.
While the former spectral component is due to the fast polymer segments of the rate
distribution, the latter is due to the slow segments. Hence, in both the $^2$H NMR
experiments, the existence of dynamical heterogeneities is demonstrated without invoking
any model assumptions. However, the present data do not provide information whether or
not homogeneous dynamics yields an additional contribution to the nonexponential
relaxation. In future work, we will show that quantification of homogeneous and
heterogeneous contributions is possible based on analysis of $^2$H NMR three-time
correlation functions, as was demonstrated for neat molecular and polymeric glass
formers.\cite{Roessler,3T}

Analysis of $^2$H NMR spin-lattice relaxation and line shape gave further evidence for
the existence of a continuous rather than a discontinuous distribution of correlation
times for all studied polymer electrolytes PPO-LiClO$_4$. Specifically, such shape of the
rate distribution is consistent with a $^2$H spin-lattice relaxation following a
stretched exponential, as was observed in the experiment. Moreover, in an analysis of the
two-phase spectra, a continuous distribution $G(\log \tau)$ was indicated by a continuous
variation of the weighting factor $W(T)$, quantifying the relative contribution of the
narrow line. However, the width of the distribution of correlation times strongly depends
on the salt concentration. While the width is moderate for both low and high salt
concentrations, an extremely broad distribution exists for the intermediate salt content
15:1 PPO-LiClO$_4$, leading to the observation of two-phase spectra over a broad
temperature range. For 15:1 PPO-LiClO$_4$, we used information about the temperature
dependent shift of $G(\log \tau)$ from our concomitant $^2$H NMR stimulated-echo
study\cite{UP} to estimate the width of this distribution from the temperature behavior
of the weighting factor $W(T)$. An estimate of more than five orders of magnitude for the
full width at half maximum shows that this polymer electrolyte exhibits very pronounced
dynamical heterogeneities.

Since it is established that presence of salt slows down the polymer dynamics, it is
reasonable to assume that the shape of the distribution of correlation times for the
polymer segmental motion reflects features of the salt distribution in polymer
electrolytes PPO-LiClO$_4$. Thus, the width of $G(\log \tau)$ implies strong fluctuations
in the local salt concentration for the intermediate 15:1 composition, while the
structural diversity is reduced for both high and low salt content. In the literature,
the possibility of liquid-liquid phase separation into salt-rich and salt-depleted
regions was controversially discussed for intermediate salt
concentrations.\cite{Vachon_1,Vachon_2,Furukawa,Bergman,Carlsson_1,Carlsson_2,Carlsson_3,Torell}
Given the rate of the polymer segmental motion is related to the local salt
concentration, the existence of well-defined microphases should manifest itself in the
existence of a bimodal distribution $G(\log \tau)$. Our $^2$H NMR results for 15:1
PPO-LiClO$_4$ indicate the existence of a continuous $G(\log \tau)$ so that they are at
variance with a bimodal distribution of correlation times and, hence, of local salt
concentrations unless the contributions from salt-rich and salt-depleted regions overlap
in large parts. On the other hand, they are in harmony with short ranged fluctuations of
the local salt concentration. In this case, one expects a large variety of local
structural motifs leading to a broad continuous $G(\log \tau)$. For example, our $^2$H
NMR findings for the intermediate salt content are consistent with a RMC model of 16:1
PPO-LiClO$_4$, which features salt-rich and salt-depleted regions on a short length scale
of about $\mathrm{10\,\AA}$,\cite{Carlsson_3} corresponding to about two monomeric units.
When we consider the cooperativity of molecular dynamics near the glass transition, the
chain connectivity and the effect that salt migration renders the spatial salt
distribution time depend on the time scale of the polymer segmental motion, it is highly
likely that a continuous distribution of correlation times $G(\log \tau)$ results from
this RMC model, consistent with our $^2$H NMR results.

\begin{acknowledgments}

We are grateful to J.\ Jacobsson for providing us with deuterated PPO. Further, we thank
H.\ Eckert for allowing us to use his NMR laboratory and S.\ Faske for helping us to
prepare the samples. Finally, funding of the Deutsche Forschungsgemeinschaft (DFG)
through the Sonderforschungsbereich 458 is gratefully acknowledged.
\end{acknowledgments}

\end{document}